\documentclass{myelsart} 
\usepackage{graphics}
\usepackage{overcite} 
\usepackage{rotating}

\begin{document}

\begin{frontmatter}

\title{A simple measure of native-state topology and chain connectivity predicts the folding rates of two-state proteins with and without crosslinks}

\author{Purushottam D.\ Dixit$^{1}$ and Thomas R.\ Weikl}
\address{Max Planck Institute of Colloids and Interfaces, Theory Department,14424 Potsdam, Germany}
\thanks{Present address: Department of Chemical Engineering, Indian Institute of Technology, 
Bombay, Powai, Mumbai 400076, India}

\begin{abstract}

The folding rates of two-state proteins have been found to correlate with simple measures of native-state topology. The most prominent among these measures is the relative contact order (CO), which is the average CO or `localness'  of all contacts in the native protein structure, divided by the chain length. Here, we test whether such measures can be generalized to capture the effect of chain crosslinks on the folding rate. Crosslinks change the chain connectivity and therefore also the localness of some of the the native contacts. These changes in localness can be taken into account by the graph-theoretical concept of effective contact order (ECO). The relative ECO, however, the natural extension of the relative CO for proteins with crosslinks, overestimates the changes in the folding rates caused by crosslinks. We suggest here a novel measure of native-state topology, the relative logCO, and its natural extension, the relative logECO. The relative logCO is the average value for the logarithm of the CO of all contacts, divided by the logarithm of the chain length. The relative log(E)CO reproduces the folding rates of a set of 26 two-state proteins without crosslinks with essentially the same high correlation coefficient as the relative CO. In addition, it also captures the folding rates of 8 two-state proteins with crosslinks.
\vspace*{-0.3cm}
\end{abstract}
\end{frontmatter}

\section{Introduction}
 
Small, single-domain proteins often are two-state folders \cite{Jackson98,Fersht99,Grantcharova01}. These proteins fold from the denatured to the native state without populating experimentally detectable intermediate states. The folding times of two-proteins have been found to vary over many orders of magnitude, from microseconds to seconds  {\cite{Jackson98,Fersht99,Grantcharova01}}. In 1998, Plaxco et al.\ \cite{Plaxco98,Plaxco00} made the remarkable observation that the the folding rates, the inverse folding times, correlate with a simple measure of native-state topology, the relative contact order (CO). Subsequently, comparable correlations have also been found for other simple measures of native-state topology such as `long-range order' \cite{Gromiha01}, the number of native contacts \cite{Makarov02,Makarov03}, the `total contact distance' \cite{Zhou02}, `cliquishness' \cite{Micheletti03}, and local secondary structure content \cite{Gong03}.

A deeper understanding of these simple topological measures requires a test of their assumptions and implications. The most prominent topological measure, the relative CO, is defined as the average CO or `localness' of all native contacts, divided by the chain length $N$. The localness or CO of a contact between residues $i$ and $j$ is the number $|i-j|$ of covalently connected residues between the two residues. The correlation between the relative CO and the folding rates of two-state proteins implies that proteins with many local contacts (e.g., $\alpha$-helical proteins) fold faster than proteins with predominantly nonlocal contacts (e.g., some $\beta$-sheet proteins) \cite{Jackson98,Fersht99,Grantcharova01}. The localness of a contact is a measure for the length of the loop that has to be closed to form the contact from the fully unfolded state. Since small loops in a flexible chain molecule on average close faster than larger loops, it seems understandable that proteins with small relative CO fold faster than proteins with larger relative CO.

In this article, we consider a simple test of the loop-closure principle underlying the relative CO. Introducing covalent chain crosslinks such as disulfide bonds into the protein chain decreases the localness of some of the native contacts, since the crosslinks `short-circuit' the chain.  The crosslinks typically lead to an increase in the folding rate \cite{Otzen98,Grantcharova00,Schoenbrunner97,Camarero01}, which is in qualitative agreement with arguments based on native-state topology. Here, we test whether topological measures of contact localness are able to reproduce these folding rate changes quantitatively.  A natural extension of the localness of a contact in a crosslinked chain is the effective contact order (ECO)\cite{Fiebig93,Dill93}, the {\em minimum} number of covalently connected residues between the two residues in contact. The ECO is a measure for the smallest loop that has to be closed to form a contact in a crosslinked, but otherwise unfolded chain. Without crosslinks, the ECO of a contact between two residues $i$ and $j$ reduces to the CO, the sequence separation $|i-j|$. 

We test and compare two topological measures based on localness. The first measure is the relative CO, and its natural extension for crosslinked chains, the relative ECO. The second measure is a novel measure, the relative logCO and its natural extension, the relative logECO. The relative logCO is defined as the average logarithm of the localness of all native contacts, devided by the logarithm of the chain length. The logarithm of the localness, i.e.\ the loop length, of a contact is an estimate for the chain entropy loss caused by the loop closure. The relative logCO and logECO therefore may be seen as naive measures of entropic folding barriers. The relative CO and relative logCO exhibit essentially the same high correlations with the folding rates of 26 two-state proteins without crosslinks. In addition, the relative logECO also captures the folding rates of 8 two-state proteins with crosslinks. The relative ECO, in contrast, seems to overestimate the folding rates of these proteins.

\section{Methods and results}

The relative ECO of a protein structure is defined as 
\begin{equation}
\mbox{rel.\ ECO} = \frac{1}{M N}\sum_{i=1}^M \mbox{ECO}(i)  
\end{equation}
The sum is taken over all contacts $i$ between non-hydrogen atoms of different residues, with total number $M$, and $N$ is the chain length, the total number of residues. { The ECO of contact $i$ is the minimum number of covalently connected residues between the residues in contact. More precisely}, the ECO is the length of the shortest path between the two residues of the contacting atoms, where each step on this path is a step between covalently connected residues. As Plaxco et al.\ \cite{Plaxco98,Plaxco00} , we define two non-hydrogen atoms to be in contact if their distance is less than 6 \AA. 

\begin{figure}
\begin{center}
\resizebox{0.9\linewidth}{!}{\includegraphics{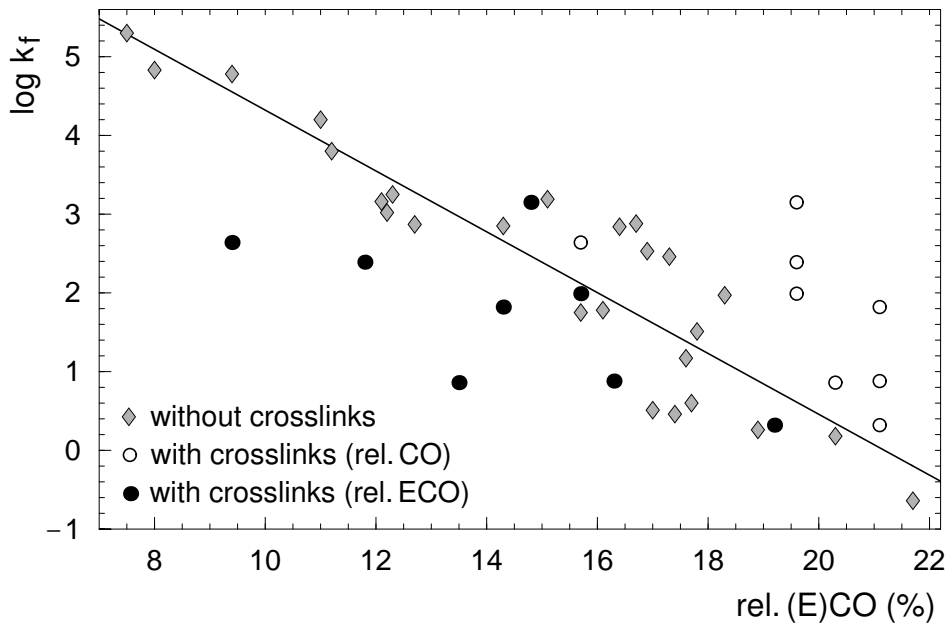}}
\end{center}
\caption{Relative CO of 26 two-state proteins without crosslinks (gray diamonds), relative CO of 8 two-state proteins with crosslinks (open circles), and relative ECO of these 8 proteins (filled circles) plotted against the decadic logarithm of their folding rates $k_f$. The regression line for the 26 proteins without crosslinks is given by $\log k_f = 8.18 - 0.386 \times (\mbox{rel.~CO})$ and provides a topology-based estimator for the folding rates of such proteins. The location of the majority of filled circles clearly below the regression indicates that the relative ECO, the natural extension of relative CO to proteins with crosslinks, tends to overestimate the folding rates of these proteins. The proteins are listed in the tables 1 and 2.
}
\label{figure_reco}
\end{figure}

For proteins without crosslinks, the relative ECO of the protein structure is identical with the relative CO. Grantcharova et al.\cite{Grantcharova01} have considered a set of 26 proteins without crosslinks, extending a previous set of Plaxco et al.\cite{Plaxco00} by two proteins. In Fig.\ \ref{figure_reco}, the relative CO of these 26 proteins is plotted against the decadic logarithm of their folding rates (gray diamonds), together with the relative CO (open circles) and the relative ECO (filled circles) of 8 two-state proteins with crosslinks. For the 26 proteins without crosslinks, the Pearson correlation coefficient between folding rate and relative CO is 0.92. The line in Fig.\ \ref{figure_reco} represents the regression line for this proteins. The position of the open circles above this regression line indicates that the relative CO of the 8 proteins with crosslinks underestimates the folding rates of these proteins. This is not unexpected, since the relative CO does not capture crosslinks, which speed up the folding process. The standard deviation of the open circles in vertical direction from the regression line is 1.42, which is significantly larger than the standard deviation of 0.61 for the 26 proteins without crosslinks. On the other hand, the relative ECO overestimates the folding rate of the proteins with crosslinks. The majority of the filled circles is located clearly below the regression line for the proteins without crosslinks, and the standard deviation of the 8 points from the regression line is 1.23. Despite the small number of data points, this deviation for the relative ECO provides a relatively clear, negative answer, since it could only be `compensated' in a much larger data set. For example, suppose we hypothetically add 8 `good' data points with the same standard deviation 0.61 as the 26 proteins without crosslinks to the 8 `poor' data points for the crosslinked proteins with standard deviation 1.23. The resulting set of 16 data points still has a standard deviation of $\sqrt{(1.23^2+0.61^2)/2}=0.97$,  which is significantly larger than the deviation 0.61 for the proteins without crosslinks.

\begin{figure}
\begin{center}
\resizebox{0.9\linewidth}{!}{\includegraphics{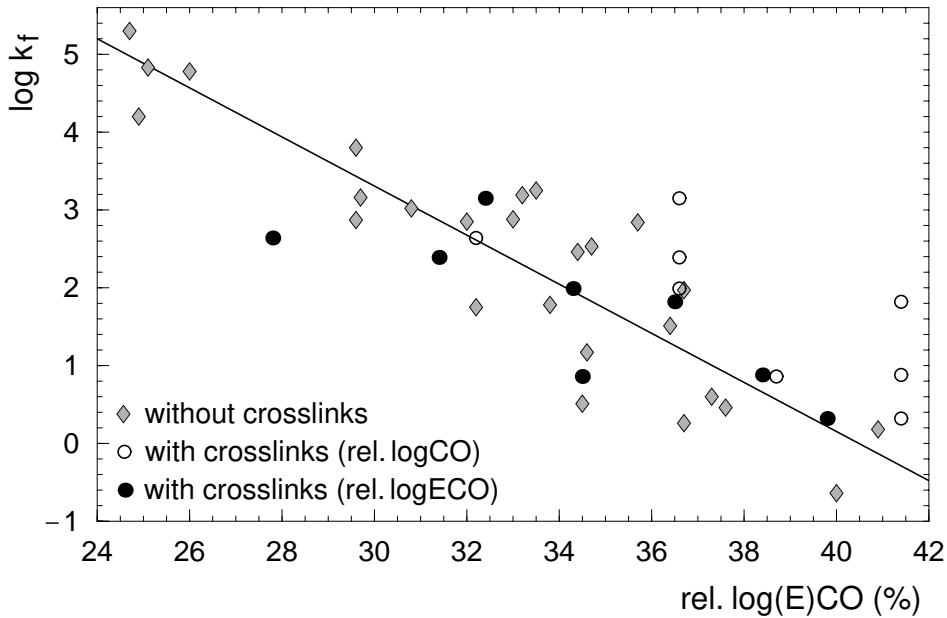}}
\end{center}
\caption{Relative logCO of 26 two-state proteins without crosslinks (gray diamonds), relative logCO of 8 two-state proteins with crosslinks (open circles), and relative logECO of these 8 proteins (filled circles) plotted against the decadic logarithm of their folding rates. The regression line for the 26 proteins without crosslinks is given by $\log k_f = 12.77 - 0.315 \times (\mbox{rel.~logCO})$. The standard deviation in vertical direction from the regression line is 0.70 for the filled circles, which is only slightly larger than the standard deviation 0.67 for the gray diamonds. This indicates that the relative logECO provides a simple, topology-based estimator for the folding rates of proteins both with and without crosslinks. In the absence of crosslinks, the relative logECO is identical with the relative logCO.}
\label{figure_rlogeco}
\end{figure}

In Fig.\ \ref{figure_rlogeco}, we consider the relative logECO, a novel measure of native-state topology and  chain connectivity,  defined as
\begin{equation}
\mbox{rel.\ logECO} = \frac{1}{M \log N}\sum_{i=1}^M \log\left[\mbox{ECO}(i)\right]  \label{rellogECO}
\end{equation}
For the 26 proteins without crosslinks, the relative logECO is identical with the relative logCO = $\sum_{i=1}^M \log\left[\mbox{CO}(i)\right]/(M \log N)$.  The relative logCO correlates with the foldings rates of these 26 proteins with a Pearson coefficient of 0.90, which is only slightly smaller than the correlation coefficient 0.92 for the relative CO.\footnote{ The small deviation between the correlation coefficients 0.90 and 0.92 is within reasonable error estimates of the coefficients. In a jack-knife approach,  these errors can be estimated by considering, for example, all subsets of the 26 data points  obtained by deleting up to two data points. For the relative ECO, the correlation coefficients of these subsets vary from 0.88 to 0.94, with a standard deviation of 0.01. For the relative logECO, the correlation coefficients of the subsets range from 0.86 to 0.93, with the same standard deviation 0.01.}
In addition, the relative logECO captures the folding rates of the 8 proteins with crosslinks. The standard deviation of the filled circles from the regression line of the 26 proteins without crosslinks is 0.70 and, thus, comparable to the standard deviation 0.67 for these 26 proteins. The relative logECO therefore provides a simple estimator for the folding rates of two-state proteins both with and without crosslinks. 

\section{Discussion and conclusions}

The correlation between folding rates and simple topological measures of two-state proteins has inspired various models of protein folding that are based on native-state topology \cite{Galzitskaya99,Munoz99,Bruscolini02,Alm99,Alm02,Debe99,Guerois00,Clementi00,Hoang00,Li01,Portman01,Kameda03,Klimov02,Flammini02,Weikl03a,Weikl03b,Weikl05}.  A deeper understanding of the remarkable success of the topological measures in reproducing the folding rates of two-state proteins requires a thorough test of the implications of these measures\cite{Wallin05}. Two-state proteins with crosslinks provide an excellent opportunity to test the `localness hypothesis' of some of the measures. The relative logECO passes this test, at least for a currently available set of 8 two-state proteins with crosslinks, and thus can be used to estimate the folding rates of two-state proteins both with and without crosslinks. 

The topological measures that have been considered here are based on physical loop-closure principles. The ECO of a contact is an estimate for the length of the loop that has to be closed to form this contact in the unfolded protein chain. For large loops, the logarithm of the loop length is proportional to the loop closure-entropy for forming this contact in the unfolded state \cite{Jacobson50,Chan90,Camacho95,Galzitskaya99,Zhou03,Zhou04}. The logarithm of the ECO in Eq.~(\ref{rellogECO}) thus can be interpreted as a loop-closure entropy.  The relative logECO is the average over the logarithm of the ECOs for all native contacts, multiplied by a prefactor $1/\log(N)$ where $N$ is the chain length. To interpret this prefactor, it is important to note that the average over the logarithm of the ECOs clearly overestimates the folding barrier. The reason is that the loop-closure cost for contacts formed late in the folding process can be reduced by contacts that have been formed earlier  \cite{Weikl03a,Weikl03b,Weikl05}. This overestimate should increase with the chain length $N$. The prefactor $1/\log(N)$ in Eq.~(\ref{rellogECO}) therefore may be seen as a heuristic, chain-length dependent correction of this overestimate, and the relative logECO as a naive estimate of entropic loop-closure barriers for folding.

Topological measures without chain-length dependent prefactors exhibit weaker correlations with the folding rates of two-state proteins. In the case of the relative CO, the prefactor is $1/N$. The related topological measure without this prefactor has been termed absolute CO \cite{Grantcharova01,Ivankov03}. For the 26 proteins without crosslinks considered here, the correlation coefficient  between absolute CO and the folding rates is 0.69, significantly smaller than the correlation coefficient 0.92 for the relative CO. The correlation coefficient for the absolute logCO = $\sum_{i=1}^M \log\left[\mbox{CO}(i)\right]/M $ is 0.80. This correlation coefficient is significantly smaller than the cofficient 0.90 for the relative logCO. 

Clearly, simply topological measure have limitations in reproducing or predicting folding rates. One of these limitations seems to be exemplified by the three src SH3 domain mutants with crosslinks listed in Table 2. The mutant with crosslink between residues 35 and 50 has the largest folding rate among the mutants. But the relative ECO and logECO of this mutant are only slightly smaller than the corresponding values for the  mutant with crosslink between residues 1 and 25, and larger than the values for the circularized mutant with crosslink between residues 1 and 56. The reason seems to be that the crosslink between residues 35 and 50 stabilizes the distal hairpin of the src SH3 domain. Mutational analysis of the wildtype src SH3 domain indicates that this $\beta$-hairpin is a central structural element in the transition state for folding \cite{Riddle99}. This seems to explain why crosslinking the hairpin has a particularly strong impact on the folding rate. The effect of native-state topology and crosslinks on the kinetics thus can also depend on structural details of transition states or native states beyond the overall localness of contacts in these states.

Another limitation is that simple, topology-based measures can't capture sequence-dependent effects. Single-residue mutations and even relatively large changes in the sequence typically have a `less than tenfold effect' \cite{Baker00} on the folding rate. These changes are comparable to the standard deviations of the folding rates from the regression lines in the correlation analysis of the topological measures. For the relative CO and logCO considered here, these standard deviations are between 0.6 and 0.7 on the decadic logarithmic scale (see above), which corresponds to an average error of a factor 5 in rate predictions. For some proteins, however, larger mutation-induced changes in the folding rate have been observed \cite{Burton96,Yang03}. On average, the effect of `topological mutations' such as the introduction or deletion of  crosslinks on the folding kinetics is significantly stronger than the effect of single-residue mutations. This is not astonishing, since these mutants change the connectivity of the protein chain, not only the local energetics.  Other examples of `topological mutants' are circular permutants in which the wild-type termini of the protein chain are connected and new termini are created by cleaving the chain somewhere else \cite{Viguera96,Otzen98,Lindberg01,Lindberg02,Miller02}. Circular permutation of the protein S6 has a drastic effect on the transition state \cite{Lindberg02},  which has been captured in a simple ECO-based model that predicts protein folding routes from native structures \cite{Weikl03b}.

 \newpage

\begin{table}

Table 1: Two-state proteins without crosslinks \\[0.1cm]
\begin{tabular}{llcccc}
protein  & PDB file & $\log(k_f)^a$  & rel.\ CO (\%) & rel.\ logCO (\%) & length (residues)  \\       
\hline
Cyt b$_{562}$             & 256B                  & 5.30 & 7.5 & 24.7 & 106 \\ 
myoglobin                    & 1BZP                  & 4.83 & 8.0 & 25.1 & 153 \\ 
$\lambda$-repressor  & 1LMB3  & 4.78 & 9.4 & 26.0 & 80 \\ 
PSBD                           & 2PDD$^b$       & 4.20 & 11.0 & 24.9 & 41 \\ 
Cyt c                             & 1HRC$^c$     & 3.80 & 11.2 & 29.6 & 104 \\ 
Im9                                & 1IMQ                   & 3.16 & 12.1 & 29.7 & 86 \\ 
ACBP                            & 2ABD                 & 2.85 & 14.3 & 32.0 & 86 \\ 
Villin 14T                      & 2VIK                    & 3.25 & 12.3 & 33.5 & 126 \\
N-term L9                     & 1DIV$^d$        & 2.87 & 12.7 & 29.6 & 56 \\
Ubiquitin                       & 1UBQ                 & 3.19 & 15.1 & 33.2 & 76 \\ 
CI2                                 & 2CI2$^e$                   & 1.75 & 15.7 & 32.2 & 64 \\
U1A                                & 1URNA    & 2.53 & 16.9 & 34.7 & 96 \\ 
Ada2h                          & 1AYE$^f$ & 2.88  & 16.7 & 33.0  & 80 \\
Protein G                      & 1PGB                  & 2.46 & 17.3 & 34.4  & 56 \\
Protein L                       & 1HZ6A$^g$ & 1.78 &  16.1 & 33.8 & 62 \\ 
FKBP                            & 1FKB                  & 0.60 & 17.7  & 37.3 & 107 \\
HPr                               & 1POH                  & 1.17 & 17.6  & 34.6 & 85 \\
MerP                             & 1AFI                    & 0.26 & 18.9  & 36.7 & 72 \\
mAcP                            & 1APS                  & $-0.64$ & 21.7 & 40.0 &   98 \\
CspB                            & 1CSP                  & 2.84  & 16.4  & 35.7  & 67 \\
TNfn3                           & 1TEN$^h$  & 0.46  & 17.4  & 37.6 & 89 \\
TI I27                            & 1TIT                    & 1.51  & 17.8 & 36.4 &  89 \\
Fyn SH3                        & 1SHF                  & 1.97 & 18.3 & 36.7 & 59 \\
Twitchin                        & 1WIT                  & 0.18  & 20.3 & 40.9 & 93 \\
PsaE                            & 1PSF                   & 0.51 & 17.0 & 34.5 & 69 \\
Sso7d                          & 1BNZA     &  3.02 & 12.2 & 30.8 & 64\\
\hline
\end{tabular}
\vspace{0.2cm}

$^a$Experimental values for the folding rates $k_f$ are from Table 1 of Grantcharova et al.\cite{Grantcharova01}. $^b$Residues 3 to 43.  $^c$Residues 1 to 104.  $^d$Residues 1 to 56.  $^e$Residues 20 to 83.  $^f$Residues 4A to 85A.  $^g$Residues 1 to 62.  $^h$Residues 803 to 891.  
-- For NMR structures with multiple models, the values for the rel.\ CO and rel.\ logCO are averages over all models. Alternate locations for atoms in PDB files have been discarded to avoid double or triple counting of corresponding contacts.  
\end{table}

\clearpage

\begin{sidewaystable}
Table 2: Two-state proteins with crosslinks \\[0.1cm]

\begin{tabular}{lccccccccc}
protein  & Ref.\ & PDB file & length & crosslink(s)  & $\log(k_f)$$^{a}$ & rel.\ ECO & rel.\ CO & rel.\ logECO & rel.\ logCO    \\  
\hline
circularized CI2 & \cite{Otzen98} & 2CI2$^b$     &   64     &      3/63          &    2.64         &    9.4   & 15.7 & 27.8          & 32.2    \\
src SH3 with SS 35/50 & \cite{Grantcharova00}  & 1SRL  & 56  &  35/50   & 3.15   & 14.8 & 19.6  & 32.4 &  36.6\\
src SH3 with SS 1/25   & \cite{Grantcharova00}  & 1SRL  & 56  & 1/25      & 1.99  &  15.7 & 19.6 & 34.3 & 36.6 \\
circularized src SH3                & \cite{Grantcharova00}  &1SRL    & 56 & 1/56     &  2.39  &   11.8 & 19.6 & 31.4 & 36.6 \\
wild-type tendamistat              & \cite{Schoenbrunner97}  & 2AIT      & 74 & 11/27, 45/73 & 1.82 & 14.3 & 21.1 & 36.5 & 41.4 \\
tendamistat w/o SS 11/27 & \cite{Schoenbrunner97}  & 2AIT     & 74  & 45/73  & 0.88 & 16.3 & 21.1 & 38.4 & 41.4 \\ 
tendamistat w/o SS 45/73 & \cite{Schoenbrunner97} & 2AIT     & 74  & 11/27  & 0.32 &19.2 & 21.1 & 39.8 & 41.4 \\
circularized c-Crk SH3 & \cite{Camarero01} & 1CKB$^c$      & 57$^d$  & 1/56$^d$    & 0.86 & 13.5 & 20.3 & 34.5 & 38.7 \\
\hline
\end{tabular}

\vspace{0.2cm}
$^a$Folding rates $k_f$ in pure water and units of $1/s$. For circularized CI2 and the crosslinked mutants of the the src SH3 domain, folding rates in water have been extrapolated using the $m_f$-values given in the references. $^b$Residues 20 to 83. $^c$Residues  135 to 190. $^d$The N- and C-termini are crosslinked via an inserted glycine residue. To calculate the rel.\ ECO and rel.\ logECO for the circularized chain, we simply assume that this glycine residue makes 25 non-hydrogen atom contacts with the nearest neighbor residues 1 and 56, and 10 contacts with the next-nearest neighbor residues 2 and 55 (these are typical numbers for glycine residues), but makes no contacts with other residues. We also assume the residues 1 and 56 have 10 contacts in the circularized chain. 
\end{sidewaystable}

\clearpage

\clearpage

\end{document}